\documentstyle [prl,aps]{revtex}
\begin{document}
\tighten
\draft
\twocolumn
\title{Filtering post-Newtonian gravitational waves  from coalescing binaries}
\author{\it B. S. Sathyaprakash}
\address {Inter-University Centre for Astronomy and Astrophysics,
Post Bag 4, Ganeshkhind Pune 411~007, India \\
E-mail: bss@iucaa.ernet.in}
\maketitle
\begin {abstract}
Gravitational waves from inspiralling binaries are expected to be detected
using a data analysis technique known as {\it matched filtering.} This
technique is applicable whenever the form of the signal is known accurately.
Though we know the form of the signal precisely, we will not know {\it a
priori} its parameters. Hence it is essential to filter the raw output
through a host of search templates each corresponding to different values
of the parameters. The number of search templates needed in detecting the
Newtonian waveform characterized by three independent parameters is itself
several thousands.  With the inclusion of post-Newtonian corrections the
inspiral waveform will have four independent parameters and this, it was
thought, would lead to an increase in the number of filters by several
orders of magnitude---an unfavorable feature since it would drastically
slow down data analysis. In this paper I show that by a judicious choice
of signal parameters we can work, even when the first post-Newtonian
corrections are included, with as many number of parameters as in the
Newtonian case. In other words I demonstrate that the effective
dimensionality of the signal parameter space does not change when first
post-Newtonian corrections are taken into account.
\end {abstract}
\pacs{PACS numbers: 04.30.+x, 04.80.+z}

Coalescing systems of compact binaries are the most promising sources
of gravitational radiation for the planned laser interferometric gravitational
wave detectors. As a binary system of  stars inspirals, due to radiation
reaction, the gravitational wave sweeps-up in amplitude and frequency.
The resulting inspiral waveform is often called the chirp waveform. As
the wave sweeps-up in frequency secular post-Newtonian corrections to the
phase of the waveform become substantial \cite {3mn}. When such corrections
are incorporated in the templates that are  used in detecting the
gravitational wave  signal it may be possible to glean useful astrophysical
information such as the masses of the component stars \cite {3mn,BS},
their equation of state, etc. In addition, observation of several such
coalescence events can facilitate an accurate determination of cosmological
parameters, like the Hubble constant and after a few hundred events
the density parameter \cite {Schutz86,Mark}. In order to extract information
of this kind, it is absolutely essential  that the parameters of the wave
form be determined very accurately.

Optimal Weiner filtering or matched filtering is a data analysis technique
that allows a very precise determination of signal parameters \cite {Th300}.
The method consists in correlating the raw output of a detector with a
waveform, variously known as a template or a filter, whose Fourier transform
is the Fourier transform of the signal divided by the noise power spectrum.
A decision about the presence or the absence of a signal is made by looking
at the maximum of the correlation.  A template whose parameters are exactly
matched with those of the signal enhances the signal-to-noise ratio (SNR)
in proportion to the square-root of the number of cycles that the signal
spends in the detector output, as opposed to the case when the shape of the
waveform is not known a priori and all that can be done is to pre-bandpass
filter the detector output to the frequency band where the signal is assumed
to lie, and then look at the SNR for each data point in the time domain
individually \cite {Schutz89}. For an interferometric detector, like the
LIGO or the VIRGO, operating with a lower frequency cutoff $\sim$~40~Hz and
an upper cutoff $\sim$~1~kHz, this means an amplification in the
SNR~$\sim$~30-40 for the inspiral waveform from a typical binary. This
enhancement in the SNR not only increases the number of detectable events
but, more importantly, it also allows a more accurate determination of
signal parameters --- the error in the estimation of a parameter being
inversely proportional to the SNR.  In order to take the full advantage
of matched filtering it is essential that the inspiralling  binary waveform,
and in particular the evolution of its phase, be known to a very high
degree of accuracy \cite {3mn}. A mismatch in the phases of the template
and the signal can drastically reduce the SNR. Though we are confident
about the event rate, and hence know the probability of the signal being
present in the data train, we will not know a priori what the signal
parameters are. Consequently, the detector output needs to be filtered
through a number of templates each corresponding to a particular set of
``test'' parameters.  The number of search templates needed to cover the
astrophysically relevant range of the parameter space depends primarily
on the dimensionality of the parameter space.

Cutler et al. \cite {3mn} pointed out that for the purpose of constructing
a set of matched filters to detect gravitational waves from inspiralling
binaries it is both necessary and sufficient to work with the so called
{\it restricted} post-Newtonian gravitational waveform. In this
approximation one incorporates the post-Newtonian corrections only to the
phase of the waveform working always with just the Newtonian amplitude.
Going beyond the restricted post-Newtonian approximation is not
expected to change appreciably the magnitude of the statistical
errors in the parameter extraction, so the restricted post-Newtonian
approximation can be used to estimate statistical errors. However,
in the post-detection analysis it is necessary to employ more accurate
templates since the use of just the restricted post-Newtonian waveform
would give rise to some systematic errors.  In the restricted
post-Newtonian approximation the gravitational waves from a binary
system of stars, modelled as point masses orbiting about each other in
a circular orbit, induce a strain $h(t)$ at the detector given by
\begin {equation}
h(t) = A (\pi f(t) )^{2/3} \cos \left [\varphi (t) \right ].
\label {wave}
\end {equation}
where $f(t)$ -- the instantaneous gravitational wave frequency -- is
equal to twice the orbital frequency of the binary; the
constant $A$ involves the distance to the binary, its reduced
and total mass, and the antenna pattern of the detector.
The detailed form of $A$ will not be of any concern in this paper.
The phase of the waveform can be schematically written as
\begin {equation}
\varphi(t) = \varphi_N(t) + \varphi_{P^1N} + \varphi_{P^{1.5}N} + \ldots.
\label {phase}
\end {equation}
Here $\varphi_N(t)$ is the dominant Newtonian part of the phase and
$\varphi_{P^nN}$ represents the $n$th order post-Newtonian correction
to it. In the quadrupole approximation we have only the Newtonian part
of the phase given by \cite {Th300}
\begin {equation}
\varphi (t) = \varphi_N (t) =
{16 \pi f_a \tau_N \over 5}
\left [ 1 - \left ({f\over f_a}\right )^{-5/3} \right] + \Phi.
\label {phaseN}
\end {equation}
Here $f(t)$ is the instantaneous Newtonian gravitational wave frequency
given implicitly by
\begin {equation}
t - t_a = \tau_N
\left [ 1 - \left ( {f \over f_a} \right )^{-8/3} \right ],
\label {frequencyN}
\end {equation}
$\tau_N$ is a constant having dimensions of time given by
\begin {equation}
\tau_N = {5 \over 256} {\cal M}^{-5/3} (\pi f_a)^{-8/3},
\label {NCT}
\end {equation}
and $f_a$ and $\Phi$ are the instantaneous gravitational wave frequency
and the phase of the signal, respectively, at $t=t_a.$ Eventhough it is
possible to invert $f$ in terms of $t$ we shall continue to work with
(\ref {frequencyN}) since it allows a straightforward interpretation of
the parameter $\tau_N$ and later of a similar post-Newtonian parameter.
We shall refer to the time elapsed starting from an epoch when the
gravitational wave frequency is $f_a$ till the epoch when it becomes
infinite, at which time the two stars would theoretically coalesce, as
the {\it chirp time} of the signal.  In the quadrupole approximation
$\tau_N$ is the chirp time.  The Newtonian part of the phase, namely
equation (\ref {phaseN}), is essentially characterized by three parameters:
(i) the {\it time of arrival} $t_a$ when the signal first becomes
{\it visible} in the detector, (ii) the {\it phase} $\Phi$ of the signal
at the time of arrival and (iii) the {\it chirp mass} ${\cal M} =
(\mu^3 M^2)^{1/5},$ where $\mu$ and $M$ are the reduced and the total mass
of the binary, respectively.  Note that at this level of approximation
the phase (as also the amplitude) depends on the masses of the two stars
only through the above combination of the individual masses.
Consequently, two coalescing binary signals of the same chirp mass but
of different sets of individual masses would be degenerate and thus exhibit
exactly the same time evolution.  This degeneracy, as we shall see below,
will be removed when post-Newtonian corrections are included in the phase
of the waveform.

How many search templates are needed to cover an interesting range of the
parameter space if we restrict ourselves to the Newtonian waveform?
Sathyaprakash and Dhurandhar \cite {SD} have made a detailed analysis of
this question and a typical number of filters they quote is about a
thousand. They have also pointed out that the present-day computer
technology is well equipped to filter the detector output online.
However, as pointed out earlier it is not enough to consider just the
Newtonian waveform. Inclusion of post-Newtonian corrections
serve dual purpose: On the one hand unless the secular post-Newtonian
corrections are included in the phase of the search templates
there would be a severe drop in the SNR.  On the other hand, and more
importantly, in order to do interesting astrophysics
with gravitational waves it is essential to remove the degeneracy
in the waveforms by including post-Newtonian corrections.

When post-Newtonian corrections are included
the parameter space of waveforms
acquires an extra dimension. It was feared that this would mean a
severe burden on data analysis: an extra dimension of the
parameter space implies that it would
be necessary to construct for each of the thousand odd Newtonian
filters a similar number of filters corresponding to the
post-Newtonian parameter. Several authors have therefore analysed the
effectiveness of a Newtonian template with ``wrong'' parameters
to detect a post-Newtonian signal \cite {BD,KKT}. However,
these authors conclude that even after allowing for a mismatch
in the parameters of a Newtonian template and a post-Newtonian signal
the SNR would reduce by about 10-20 \%. Such a drop in the
SNR is unfavorable considering the low event rate of these
sources.  In this paper I will show that when
the first post-Newtonian corrections are included in the phase of
the waveform it is possible to make a judicious choice of the
parameters so that the parameter space essentially remains only
three dimensional. While such a strategy
is suitable for an unambiguous and easy detection it does by
no means guarantee a precise estimation of all of the binary's
parameters. The bottomline of this paper is
that the algorithm presented here enables
an enhancement in the SNR by including post-Newtonian corrections
in the search templates without at the same time causing any
extra burden on data analysis. It should however be noted that
further off-line analysis would be necessary to extract
useful astrophysical information.

With the inclusion of first post-Newtonian correction the phase of the
waveform becomes \cite {WW,CF}
\begin {equation}
\varphi (t) = \varphi_N (t) + \varphi_{P^1N} (t)
\label {phasetotal}
\end {equation}
where $\varphi_N(t)$ is given by (\ref {phaseN}) and
\begin {equation}
\varphi_{P^1N} (t) =
4 \pi f_a \tau_{P^1N} \left [ 1 - \left ( {f\over f_a} \right )^{-1} \right ].
\label {phase1PN}
\end {equation}
Now $f(t)$ is the instantaneous post-Newtonian frequency
given implicitly by
\begin {equation}
t - t_a =
\tau_N \left [ 1 - \left ( {f \over f_a} \right )^{-8/3} \right ] +
\tau_{P^1N} \left [ 1 - \left ( {f \over f_a} \right )^{-2} \right ],
\label {frequencyPN}
\end {equation}
$\tau_N$ is given by (\ref {NCT}) and $\tau_{P^1N}$ is a constant
having dimensions of time given by
\begin {equation}
\tau_{P^1N} = {5 (743 + 924 \eta) \over 64512\mu (\pi f_a)^2}
\end {equation}
The phase (\ref {phase1PN})
now contains the reduced mass $\mu$ and the parameter $\eta = \mu / M,$
in addition to the chirp mass $\cal M.$
Taking ($\cal M$, $\eta$) to be the post-Newtonian mass parameters
the total mass and the reduced mass are given by
$M = {\cal M} \eta^{-3/5},$ $\mu = {\cal M} \eta^{2/5}.$
Note that the total chirp time $\tau$ of the signal
has a Newtonian contribution $\tau_N$ and a post-Newtonian
contribution $\tau_{P^1N}:$ The time left starting from an
epoch when the gravitational wave frequency is $f_a$ until
an epoch when the frequency becomes infinite is
$\tau = \tau_N + \tau_{P^1N}.$
We shall refer to $\tau_N$ as the Newtonian chirp time and
to $\tau_{P^1N}$ as the post-Newtonian chirp time.
Note that instead of working with the parameters
${\cal M}$ and $\eta$ we can equivalently work with the
parameters $\tau_N$ and $\tau_{P^1N}.$

Thus the post-Newtonian filter is characterized by
four parameters: $\lambda_k= \{t_a,$ $\Phi,$ $\tau_N,$
$\tau_{P^1N}\}$ where we have used the symbol $\lambda_k,$ $k=1,\ldots, 4,$
to collectively denote the four parameters. Note that $t_a$ and
$\Phi$ are {\it kinematical} parameters that fix the origin of the
measurement of time and phase, respectively, whereas the Newtonian and the
post-Newtonian chirp times
are {\it dynamical} parameters in the sense that they decide the
evolution of the phase and the amplitude of the signal.
We shall now set out to see if it is possible,
for the purpose of filtering, to reduce the
dimensionality of the parameter space from four to three.

We begin by defining the scalar product of waveforms which
plays a crucial role in deciding the filters that are required
to span the range of parameters and hence to assess the
effective dimensionality of the parameter space. Given
two waveforms $g(t)$ and $h(t)$ their scalar product is defined by
\begin {equation}
\left < g , h \right > \equiv \int_{-\infty}^\infty
{\tilde g(f) \tilde h^*(f)\over S_n(f)} d f
\label {scalarproduct}
\end {equation}
where $S_n(f)$ is the {\it two-sided} detector noise power spectral density
and
$\tilde g(f) = \int^\infty_{-\infty} g(t) \exp (2\pi ift)dt$ and
$\tilde h(f) = \int^\infty_{-\infty} h(t) \exp (2\pi ift)dt,$
are the Fourier transforms of the waveforms $g(t)$ and $h(t),$ respectively.
The SNR $\rho$ obtained for a signal $h(t)$ using an optimal Weiner filter
is simply the norm of the signal computed using the above definition
of the scalar product:
\begin {equation}
\rho = \left < h, h \right >.
\label {SNR}
\end {equation}
A waveform is said to be {\it normalized} if its norm is equal to unity.
Let us consider the behavior of the scalar product of two chirp waveforms
$g(t; \lambda_k)$ and $h (t; \lambda_k^0)$ which differ in all their
parameter values, i.e., $\lambda_k$ being in general different from
$\lambda_k^0,$ and are normalized, i.e.,
$\left <h, h\right >= \left <g, g\right > = 1.$
Their scalar product $C (\lambda_k, \lambda_k^0)$ is given by
\begin {equation}
C (\lambda_k, \lambda_k^0) = \left <g (\lambda_k), h (\lambda_k^0) \right >.
\label {defcorr}
\end {equation}
Here $\lambda_k$ can be thought of
as the parameters of a signal while $\lambda_k^0$ those of
a template. Then $C (\lambda_k, \lambda_k^0)$ is the SNR obtained using
a template that is not necessarily matched on to the signal.
Since the waveforms are of unit norm
$C (\lambda_k, \lambda_k^0) = 1,$ if $\lambda_k = \lambda^0_k$ and
$C (\lambda_k, \lambda_k^0) < 1,$ if $\lambda_k \ne \lambda^0_k.$
In general, as indicated by its arguments $C (\lambda_k, \lambda_k^0)$
depends on the individual values of the parameters
both of the signal and the template. In what follows I will first
show that the SNR (\ref {defcorr}) depends only on the difference
in the parameter values $\lambda_k-\lambda_k^0.$
Secondly, I will show that a template of a given total chirp time
obtains roughly the same SNR for all waveforms of the same total chirp time
though their Newtonian and post-Newtonian chirp times may be
different from that of the template (see, however, the discussion
at the end of the paper). The former of these two results implies
uniformity in the spacing of filters \cite {SD,DS} while the latter
result facilitates a massive reduction in the number of
templates required in spanning the parameter space since instead
of constructing filters separately for each of the
Newtonian and post-Newtonian chirp times we can construct filters
simply for the total chirp time.

In the stationary phase approximation
the Fourier transform of the restricted first-post-Newtonian
chirp waveform for positive frequencies is given by \cite {Th300,SD,FC}
\begin {mathletters}
\begin {equation}
\tilde h (f) = \tilde A f^{-7/6} \exp \left [i\sum_{k=1}^4\psi_k(f)\lambda_k
- i {\pi \over 4} \right ]
\label {FT}
\end {equation}
where $\tilde A$ is a constant and
\begin {eqnarray}
\psi_1 & = & 2\pi f, \\
\psi_2 & = & 1, \\
\psi_3 & = & 2 \pi f  -{ 16 \pi f_a \over 5}+ {6\pi f_a \over 5}
\left ( {f\over f_a} \right )^{-5/3},\\
\psi_4 & = & 2\pi f - \pi f_a +
2\pi f_a \left ({ f\over f_a} \right)^{-1}.
\label {FTphase}
\end {eqnarray}
\end {mathletters}
For $f<0$ the Fourier transform is computed using the identity $\tilde
h(-f) = \tilde h^*(f)$ obeyed by real functions $h(t).$
With the above expression for the Fourier transform
the SNR (\ref {defcorr}), using (\ref {scalarproduct}), takes the form
\begin {equation}
C (\delta\lambda_k) \propto \int_0^\infty
{f^{-7/3}\over S_n(f)} \cos
\left [ \sum_{k=1}^4\psi_k (f) \delta\lambda_k \right ] df
\label {corr1}
\end {equation}
where $\delta \lambda_k = \lambda_k - \lambda_k^0.$ As in the
Newtonian case the SNR is independent of the individual parameter
values of the signal and the template: {\it For all signal--template
pairs that have the same differences in times of arrival, phases,
and chirp times one obtains the same SNR.}
Consequently, constancy of the distance,
measured using the scalar product (\ref {scalarproduct}),
between two nearest neighbour filters,
that is required in making a choice of filters, translates into the
constancy of the {\it distance}, measured using the difference
in their parameter values.

We now seek to analyse the behavior of the SNR
(\ref {corr1}) when the template's parameters are mismatched with
those of the signal. While it is essential to do the analysis
for noise power spectral density in real laser interferometers, such
as the one discussed by Finn and Chernoff \cite {FC}, the results obtained
in that case are qualitatively the same as in the case of white noise
\cite {RSJ}. In order not to divert attention from the main theme of
discussion, here I will only quote the results for white noise.
$C(\delta\lambda_k)$ traces out
a four-dimensional surface as we vary $\delta\lambda_k.$
In what follows I consider the two-dimensional subspace obtained
by maximizing $C (\delta\lambda_k),$ over $t_a$ and $\Phi,$ for every
pair of $\tau_N$ and $\tau_{P^1N}$ of the signal keeping the parameters
of the template $\tau_N^0$ and $\tau_{P^1N}^0$ constant.
The surface so obtained is
plotted for white noise [i.e. $S_n(f) = $~const.]
in Fig.\ref {corrsurface}. Since the parameters of the template are
constant I have shown on the x- and y-axis of Fig. \ref {corrsurface}
(and Fig.\ref {contour}) the parameters of the signal $\tau_N$ and
$\tau_{P^1N}$ and not the difference $\delta\tau_N$ and $\delta\tau_{P^1N}.$
The same surface is obtained irrespective
of what parameters we choose for the parameters of the template provided we
keep the range of the signal parameters the same. In this sense, the
correlation surface in Fig. \ref{corrsurface}
only depends on the difference in the parameters of the signal
and the template and not on their absolute values.
For the astrophysically relevant range of the masses of the two stars
(say, $M_1, M_2 \in [1, 10]~M_\odot)$
and for $f_a=100$~Hz, $\tau_N \in [4, 0.08]$~s and
$\tau_{P^1N}\in [0.3, 0.03]$~s.
In Fig.\ref{corrsurface} the post-Newtonian chirp time is varied over
the whole of its relevant range while the Newtonian chirp time
is only varied over a portion of its relevant range. The contours
of this surface shown in Fig.\ref {contour} are {\it almost}
straightlines $ \tau = \tau_N+\tau_{P^1N} = $~const., except
for the inner most one or two contours. The value of $C$ corresponding
to the inner most contour is 0.9 and reduces by 0.1 with successive
outer contours.

A useful interpretation of the surface in Fig.\ref {corrsurface}
is the following: Imagine that we have a template with parameters
corresponding to the center of the grid. It obtains an SNR of unity
with a signal whose parameters are exactly matched on to it. The
SNR that it obtains for other waveforms
is in general less than unity but, as is evident from
Fig.\ref {corrsurface}, the SNR is almost equal
to unity for every waveform whose total chirp time is the same as its
own total chirp time. Thus, if we choose our search templates
along the curve perpendicular to the contours of Fig.\ref {contour} then
we will in effect be covering the entire subspace of the signal. In
other words {\it we can span the two-dimensional $(\tau_N, \tau_{P^1N})$
subspace of the four-dimensional parameter space with just one parameter.}
This curve is an appropriate one of a family of straight lines
$\tau_N=\tau_{P^1N}+$~const.
Consequently, as far as the choice of filters is considered we need only
work with three parameters, namely $\{t_a, \Phi, \tau\}.$ This reduction
in the effective number of parameters can be traced to the fact that
there is a strong covariance between the
parameters $\tau_N$ and $\tau_{P^1N}.$ This result, together with
the details of a Monte Carlo simulation demonstrating the effectiveness
of the claim made in this paper and the results of including
higher-order post-Newtonian corrections, and noise envisaged
in real interferometers will be published elsewhere
\cite {RSJ}.

Let me conclude by making two cautionary remarks. The first one
concerns the choice of parameters: It should not be thought that
the effective dimensionality of the parameter space
is three only when the set $\{t_a, \Phi, \tau_N, \tau_{P^1N}\}$ is
employed in constructing a lattice of filters. Afterall the
reduction of dimensionality is related to the property of the
scalar product (\ref {scalarproduct}) which is re-parametrization invariant.
The advantage of the set used in this paper is that it
allows us to conclude about the
effective dimensionality without recourse to complicated mathematical
analysis. However, the final justification has to come
from a more rigorous analysis which will be taken up in a future paper
\cite {RSJ}.
The second comment is about the scope of the reduced dimensionality
of the parameter space itself: The parameter space would
be truely three-dimensional provided the correlation function is a constant
in the direction $\tau_N+\tau_{P^1N}=$~const. However, as can be
inferred from the Fig. \ref {corrsurface} and \ref {contour}, strictly
speaking, this is not the case:
The correlation function slowly decreases as we move away from the
maximum of the correlation function in the direction
$\tau_N+\tau_{P^1N}=$~const.
This means two things: (i) The argument about the reduction
in dimensionality is only valid as long as the correlation function
has not dropped too much (ideally less than about 1\%)
along lines of constant $\tau_N+\tau_{P^1N}$ and
(ii) A post-Newtonian filter of a given total chirp time
{\it cannot} be replaced by a Newtonian filter of the same chirp time.
In other words, the presence of the post-Newtonian term cannot be
mimicked by a Newtonian filter alone. However, for the astrophysically
relevant range of the parameter $\tau_{P^1N}$ it turns out that
we need only use a three-dimensional lattice of filters or at worst
two sets of a three-dimensional lattice.

I am indebted to the referee for pointing out several flaws in an
earlier version of this paper.  It is a pleasure to thank the members
of the gravitational wave group at IUCAA, especially R. Balasubramanian,
Sanjeev Dhurandhar and Kanti Jotania, for many useful discussions.

\begin {figure}
\vspace {2.8 true in}
\caption {Surface showing the maximum, over $t_a$ and $\Phi,$
of the SNR $C(\delta\lambda_k).$}
\label {corrsurface}
\end {figure}

\begin {figure}
\vspace {2.8 true in}
\caption {Contours of the SNR surface shown in Fig.1. The contours,
 are approximately straightlines $\tau_N+\tau_{P^1N}=$~const.}
\label {contour}
\end {figure}

\end {document}